# Kinematics and Velocity Ellipsoid of Halo Red Giants

Mohamed I. Nouh[1, 2] and Waleed H. Elsanhoury[2, 3]

[1]Physics Dept., College of Science, Northern Border University, Arar, Saudi Arabia.
Email: nouh@nbu.edu.sa; abdo_nouh@hotmail.com
[2]Astronomy Dept., National Research Institute of Astronomy and Geophysics (NRIAG), 11421, Helwan, Cairo, Egypt (Affiliation ID: 60030681)
[3]Physics Dept., Faculty of Science and Arts, Northern Border University, Rafha Branch, Saudi Arabia.
Email: elsanhoury@nbu.edu.sa; welsanoury@gmail.com

**Abstract**: In the present paper, we aim to determine the kinematical properties, velocity ellipsoid, and Oort constants using a sample of halo red giants. The study is based mainly on the space and radial velocities of about 1583 red giant stars collected from the SEGUE-1 and SEGUE-2 surveys. We divided the sample into three subsamples, the inner halo, the outer halo and the stars near the galactic plane. The fitting of the radial velocity equation gives a the mean of Oort constants, A = 15.6 ± 1.6 km s$^{-1}$ kpc$^{-1}$ and B= -13.9 ± 1.8 km s$^{-1}$ kpc$^{-1}$, the angular velocity |A-B|= 29.5±0.2 km s$^{-1}$ kpc$^{-1}$ implying a rotational velocity of 221.25 ± 26.66 km s$^{-1}$ if we take the distance to the Galactic center as 7.5 kpc.

1. Introduction

Stellar kinematics is an essential ingredient in the study of galactic structure and evolution. The halo stars, in particular, may be exploited to probe the formative phases of our galaxy [1].

There are a number of studies provided evidence that the halo of the Milky Way may not comprise a single population, primarily from the analysis of spatial profiles (or inferred spatial profiles) of halo objects ([2], [3]; [4]; [5]; [6]; [7]. A recent example of such an analysis is the observation of two different spatial density profiles for the distinct Oosterhoof classes of RR Lyrae variable stars in the halo [8]. In addition, tentative claims for a net retrograde motion of halo objects by previous authors supports the existence of a likely dual-component halo, [9]; [10]; [11]; [12]; [13]; [14]; [15]. Using astrophysical simulations, the galactic halo has been divided into two components, the inner halo and the outer halo [16]. The inner halo is dominated by stars which formed within the galaxy, where the outer halo is mainly composed of stars accreted through merger events.

The importance of red giant stars comes from that, they are the most luminous ones found in a population of old stars, and so are particularly useful to study the early history of the Milky Way. So, researchers use these stars like fossils, because in many cases their chemistry and motions have been unchanged since they were formed more than 10 Gyr ago. According to the Sloan Digital Sky survey's SEGUE project, there are over 5,000 giant stars, some of them as far away as 100 kiloparsecs (kpc; for comparison, the Milky Way's brightest satellite companion galaxies, the Magellanic Clouds, are only 50 kpc away). In the present paper, we are going to calculate the kinematical parameters and the rotational constants for a sample of halo red giants.

The structure of the paper is as follows: Section 2 describes the observational data. Section 3 is devoted to the calculations of the kinematical parameters of the sample. The galactic rotational constants are determined in section 4. The conclusion is given in Section 5.

2. **Observational Data**

The sample of halo red giant stars in the halo fields is selected by [17] from the SEGUE-1 and SEGUE-2 surveys, [18]; [19] and SDSS-III/SEGUE-2 ([20]; [21]) surveys, respectively. Both SEGUE surveys were spectroscopic extensions of SDSS, with the goal of acquiring broad wavelength-coverage, moderate-resolution ($R \sim 2000$) optical spectra of stars in specific Galactic populations. Carolo [22] used proper motions in combination with distance estimates and radial velocities, provide the information required to calculate the full space motions (the components of which are referred to as $U$, $V$, $W$) of our program stars with respect to the local standard of rest (LSR).

We retrieved the data with complete records of space velocities, radial velocity, proper motion, distance, and metallicity. In this context, we get 1444 stars. The effective temperature of the program stars ranges from 4266 $K^0$ to 6330 $K^0$, the metallicity ranges from -2.29 to -0.69 and distances up to 40 kpc from the Sun. Carollo [15] applied the corrections for the motion of the Sun with respect to the LSR during the course of the calculation of the full space motions; where they adopt the values of Mihalas & Binney (1981) ($U_\odot$, $V_\odot$, $W_\odot$) = (-9, 12, 7) km s$^{-1}$.

We follow the procedure introduced by [15] and divided our sample into three small samples, i.e. an inner halo of stars having distances $d \leq 15$ kpc, an outer halo for stars having distances $d = 15 - 20$ kpc and add another small sample for the stars near the galactic plane; $7 \leq R \leq 10$ kpc. Figure 1 displays the distribution of the radial velocities of the program stars with Galactic longitudes. In Figure 2 we plot the space velocities $U$ $V$ and $W$ as a function of the metallicity.

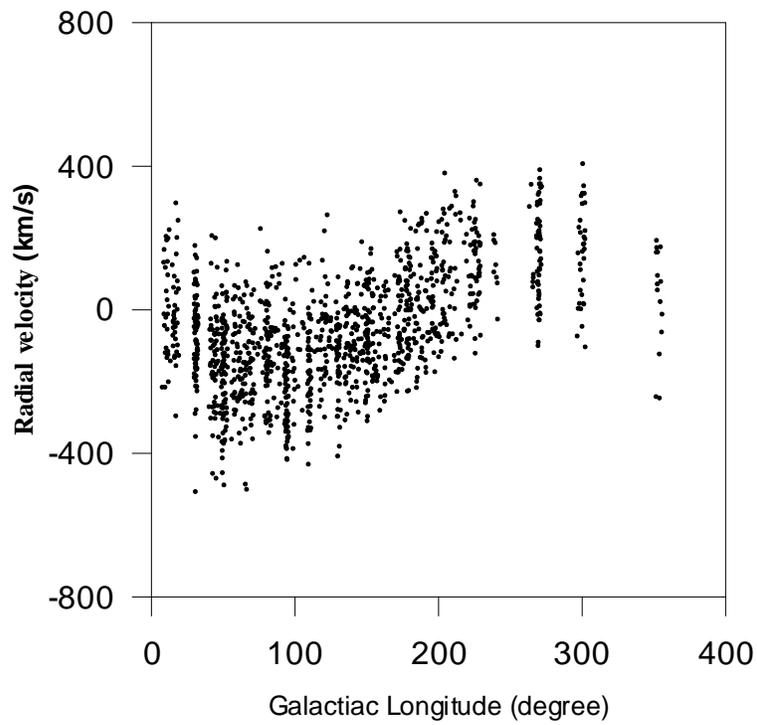

Figure 1. Distribution of observed radial velocities of the red giant stars along the Galactic longitude

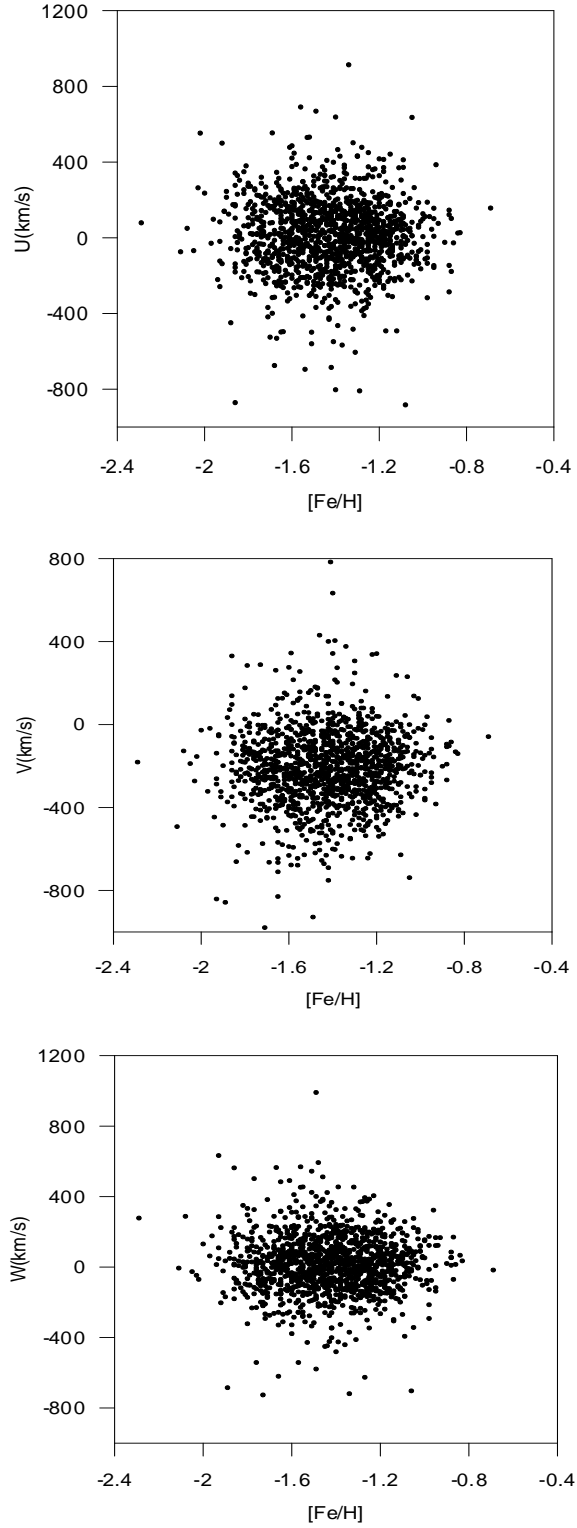

Figure 2. The U, V, and W components of the space motions of stars in our sample as a function of [Fe/H].

## 3. The Kinematical Model

We follow the computational algorithm by [23] to compute the velocity ellipsoid parameters VEPs for the above sample of data as well as the Solar elements. A brief explanation of the algorithm will be given here. The coordinates of the $i^{th.}$ star with respect to axes parallel to the original axes, but shifted to the center of the distribution, i.e. into the point $\overline{U}, \overline{V}$ and $\overline{W}$, will be $(U_i - \overline{U}); (V_i - \overline{V}); (W_i - \overline{W})$, where $U, V$ and $W$ are the components of the space velocities and $\overline{U}, \overline{V}$ and $\overline{W}$ are the mean velocities defined as:

$$\overline{U} = \frac{1}{N} \sum_{i=1}^{N} U_i; \overline{V} = \frac{1}{N} \sum_{i=1}^{N} V_i; \overline{W} = \frac{1}{N} \sum_{i=1}^{N} W_i \tag{1}$$

$N$ being the total number of the stars, and

$$U = -0.0518807421 V_x - 0.8722226427 V_y - 0.4863497200 V_z, \tag{2}$$

$$V = 0.4846922369 V_x - 0.4477920852 V_y + 0.7513692061 V_z, \tag{3}$$

$$W = -0.8731447899 V_x - 0.1967483417 V_y + 0.4459913295 V_z. \tag{4}$$

Let $\xi$ be an arbitrary axis, its zero point coincident with the center of the distribution and let $l, m$ and $n$ be the direction cosines of the axis with respect to the shifted one, then the coordinates $Q_i$ of the point $i$, with respect to the $\xi$ axis is given by:

$$Q_i = l(U_i - \overline{U}) + m(V_i - \overline{V}) + n(W_i - \overline{W}). \tag{5}$$

Let us adopt, as the measured of the scatter components $Q_i$, a generalization of the mean square deviation, defined by

$$\sigma^2 = \frac{1}{N} \sum_{i=1}^{N} Q_i^2 \tag{6}$$

From Equations (1), (5) and (6) we deduce after some calculations that

$$\sigma^2 = \underline{x}^T B \underline{x} \tag{10}$$

where $\underline{x}$ is the $(3\times 1)$ direction cosines vector and $B$ is $(3\times 3)$ symmetric matrix $\mu_{ij}$, with elements $\mu_{ij}$:

$$\left.\begin{aligned}\mu_{11} &= \frac{1}{N}\sum_{i=1}^{N} U_i^2 - \left(\overline{U}\right)^2; \quad \mu_{12} = \frac{1}{N}\sum_{i=1}^{N} U_i V_i - \overline{U}\,\overline{V}; \\ \mu_{13} &= \frac{1}{N}\sum_{i=1}^{N} U_i W_i - \overline{U}\,\overline{W}; \quad \mu_{22} = \frac{1}{N}\sum_{i=1}^{N} V_i^2 - \left(\overline{V}\right)^2; \\ \mu_{23} &= \frac{1}{N}\sum_{i=1}^{N} V_i W_i - \overline{V}\,\overline{W}; \quad \mu_{33} = \frac{1}{N}\sum_{i=1}^{N} W_i^2 - \left(\overline{W}\right)^2. \end{aligned}\right\} \tag{11}$$

The necessary conditions for an extremum are now

$$(B - \lambda I)\underline{x} = 0 \tag{12}$$

These are three homogenous equations in three unknowns, which have a nontrivial solution if and only if

$$D(\lambda) = |B - \lambda I| = 0, \tag{13}$$

where $\lambda$ is the eigenvalue, and $\underline{x}$ and $B$ are given as:

$$\underline{x} = \begin{bmatrix} l \\ m \\ n \end{bmatrix} \text{ and } B = \begin{vmatrix} \mu_{11} & \mu_{12} & \mu_{13} \\ \mu_{12} & \mu_{22} & \mu_{23} \\ \mu_{13} & \mu_{23} & \mu_{33} \end{vmatrix}$$

Equation (10) is characteristic equation for the matrix $B$. The required roots (i.e. eigenvalues) are

$$\left.\begin{array}{l}\lambda_1 = 2\rho^{\frac{1}{3}}\cos\dfrac{\phi}{3} - \dfrac{k_1}{3}; \\[6pt] \lambda_2 = -\rho^{\frac{1}{3}}\left\{\cos\dfrac{\phi}{3} + \sqrt{3}\sin\dfrac{\phi}{3}\right\} - \dfrac{k_1}{3}; \\[6pt] \lambda_3 = -\rho^{\frac{1}{3}}\left\{\cos\dfrac{\phi}{3} - \sqrt{3}\sin\dfrac{\phi}{3}\right\} - \dfrac{k_1}{3}.\end{array}\right\} \qquad (14)$$

where

$$\left.\begin{array}{l}k_1 = -(\mu_{11} + \mu_{22} + \mu_{33}), \\[4pt] k_2 = \mu_{11}\mu_{22} + \mu_{11}\mu_{33} + \mu_{22}\mu_{33} - (\mu_{12}^2 + \mu_{13}^2 + \mu_{23}^2), \\[4pt] k_3 = \mu_{12}^2\mu_{33} + \mu_{13}^2\mu_{22} + \mu_{23}^2\mu_{11} - \mu_{11}\mu_{22}\mu_{33} - 2\mu_{12}\mu_{13}\mu_{23}.\end{array}\right\} \qquad (15)$$

$$q = \frac{1}{3}k_2 - \frac{1}{9}k_1^2 \quad ; \quad r = \frac{1}{6}(k_1 k_2 - 3k_3) - \frac{1}{27}k_1^3 \qquad (16)$$

$$\rho = \sqrt{-q^3} \qquad (17)$$

$$x = \rho^2 - r^2 \qquad (18)$$

and

$$\phi = \tan^{-1}\left(\frac{\sqrt{x}}{r}\right) \qquad (19)$$

Depending on the matrix that control the eigenvalue problem [Equation (9)] for the velocity ellipsoid, we establish analytical expressions of some parameters for the correlations studies in terms of the matrix elements $\mu_{ij}$ of the eigenvalue problem for the velocity ellipsoid. The velocity dispersions $\sigma_i$ ; $i = 1, 2, 3$ could be given by

$$\sigma_i = \sqrt{\lambda_i}. \qquad (20)$$

The center of the cluster can be derived by the simple method of finding the equatorial coordinates of the center of mass for the number $N_i$ of discrete objects, i.e.

$$x_c = \left[\sum_{i=1}^{N} r_i \cos\alpha_i \cos\delta_i\right]/N,$$

$$y_c = \left[\sum_{i=1}^{N} r_i \sin\alpha_i \cos\delta_i\right]/N, \qquad (21)$$

$$z_c = \left[\sum_{i=1}^{N} r_i \sin\delta_i\right]/N.$$

The Solar motion can be defined as the absolute value of the Sun's velocity relative to the group of stars under consideration,

$$S_\odot = \left(\overline{U}^2 + \overline{V}^2 + \overline{W}^2\right)^{1/2}. \ km\,s^{-1}. \qquad (22)$$

The galactic longitude $(l_A)$ and galactic latitude $(b_A)$ of the Solar apex are

$$l_A = tan^{-1}\left(-\overline{V}/\overline{U}\right), \qquad (23)$$

$$b_A = sin^{-1}\left(-\overline{W}/S_\odot\right), \qquad (24)$$

These three parameters may be called elements of solar motion with respect to a group under consideration.

      We computed the kinematical parameters and the Solar motion for the three subsamples; the inner halo, the outer halo, and the stars near the galactic plane. The results are listed in Table 1, where row 1 is the total number of stars in each class, row 2 is the effective temperature of stars, rows 3, 4, and 5 are the average space velocities due to Galactic coordinates, rows 6, 7, and 8, are the eigenvalues, rows 9, 10, and 11 are devoted to dispersion velocities, rows 12, 13, and 14 are the direction cosines, and rows 15, 16, and 16 gives the Solar elements.

**Table 1:** Velocity ellipsoid and Solar velocity for the three subsamples.

| Parameters | Inner $d \leq 15$ kpc 926 stars | Outer $d = 15-20$ kpc 518 stars | Galactic plane $10 \geq R \geq 7$ kpc 160 stars |
|---|---|---|---|
| $\overline{U}$ $(km/s)$ | $15.05 \pm 3.88$ | $-7.19 \pm 2.68$ | $22.45 \pm 4.74$ |
| $\overline{V}$ $(km/s)$ | $-212.20 \pm 14.57$ | $-208.93 \pm 14.45$ | $-210.36 \pm 14.50$ |
| $\overline{W}$ $(km/s)$ | $16.28 \pm 4.03$ | $21.31 \pm 4.62$ | $32.93 \pm 5.74$ |
| $\lambda_1$ $(km/s)$ | 70681.7 | 81799.7 | 79029.4 |
| $\lambda_2$ $(km/s)$ | 35148.0 | 47588.5 | 36182.6 |
| $\lambda_3$ $(km/s)$ | 19123.3 | 30402.6 | 19373.3 |
| $\sigma_1$ $(km/s)$ | 265.86 | 286.01 | 281.12 |
| $\sigma_2$ $(km/s)$ | 187.48 | 218.15 | 190.22 |
| $\sigma_3$ $(km/s)$ | 138.29 | 174.36 | 139.19 |
| $(l_1, m_1, n_1)_{deg}$ | 0.056, -1.00, 0.086 | 0.00, -0.990, 0.144 | 0.103, -0.992, -0.070 |
| $(l_2, m_2, n_2)_{deg}$ | -0.92, -0.085, -0.390 | -0.906, -0.062, -0.419 | -0.697, -0.021, -0.717 |
| $(l_3, m_3, n_3)_{deg}$ | 0.40, -0.058, -0.917 | 0.423, -0.130, -0.897 | 0.710, -0.123, -0.693 |
| $S_\odot$ $(km/s)$ | $213.36 \pm 14.61$ | $210.14 \pm 14.50$ | $214.10 \pm 14.63$ |
| $l_A$ | 85.94 | -88.03 | 83.91 |
| $b_A$ | -4.37 | -5.82 | -8.85 |

The results in Table 1 show that the velocity dispersions ($\sigma_1$, $\sigma_2$, $\sigma_3$) obey the inequalities $\sigma_1 > \sigma_2 > \sigma_3$ and they behave the radially elongated velocity ellipsoid appeared, Chiba & Beers [24] obtained the same behavior in terms of the halo's kinematics. The longitude of the vertex of the velocity ellipsoid ($l_A^o$) calculated for our sample indicates that the principal axis points in the direction of the galactic center.

## 4. The Galactic Rotation Constants

The Oort constants can be related to circular velocity and thus the potential of the Galaxy in an axisymmetric approximation [25]. The first proof of the existence of the differential galactic rotation was by Oort [26, 27]. Since this time, there are several calculations for the two Oort's constants *A* and *B*.

To determine the rotation constant *A*, we follow two methods. The first, by using the fact that the radial velocity ($V_r$) shows a double sine-wave variation with the galactic longitude with an amplitude that increases linearly with distance as [28]

$$V_r = -2A (R - R_0) \sin l \cos b + K,  \quad (25)$$

where *l* and *b* are the longitude and latitude of the individual star, $R_0$ is the distance from the Sun to the galactic center and *K* is a term which can be interpreted as systematic motions of large stellar groupings, systematic errors in the radial velocities due to such causes as gravitational redshift, motions within stellar atmospheres and erroneous wave-length systems [29].

The radial distance of the star from the galactic center *R* is given by

$$R^2 = R_0^2 + d^2 - 2 R_0 \, d \cos l. \quad (26)$$

We calculated the Oort constant *A* for the three subsamples, the inner halo, the outer halo, and the stars near the galactic plane. The results are listed in Table 2, where column 1 is the first Oort constant computed from the least squares fit to Equation (25), column 2 is the *K* term, column 3 is the second Oort constant computed using the relation

$$(\sigma_2/\sigma_1)^2 = -B/(A-B), \text{ [30]}.$$

**Table 2:** Rotation constants for the three subsamples.

| Parameters | *A* (km s$^{-1}$ kpc$^{-1}$) | K-term (km s$^{-1}$) | *B* (km s$^{-1}$ kpc$^{-1}$) | $\sigma_2/\sigma_1$ |
|---|---|---|---|---|
| Galactic Plane | 16.723 ± 1.81 | -2.30 ± 0.37 | -13.610 | 0.68 |
| Inner halo | 14.592 ± 1.76 | -2.861 ± 0.37 | -14.360 | 0.70 |
| Outer halo | 14.930 ± 1.82 | -2.78 ± 0.37 | -20.420 | 0.76 |

The Oort constants can be connected to the local angular velocity through the relation $|A-B|$. According to the present result $|A-B|$ equals $29.5 \pm 0.2$ km s$^{-1}$ kpc$^{-1}$. This result agrees with [31] for the red giants ($|A-B| = 29.6 \pm 1$ km s$^{-1}$ kpc$^{-1}$) but differs from the results by previous works listed in Table (3). Table (3) lists the Oort constants calculated by different authors. The rotational velocities in column 4 are calculated assuming $R_0 = 7.5$ kpc. The negative K-terms for the three program stars not significantly different from zero. These values differ from the finding by many authors for early-type stars, they showed significant values of the K-term.

Table 3: Adopted Oort constants and comparison with previous works.

| Origin | A km s$^{-1}$ kpc$^{-1}$ | B km s$^{-1}$ kpc$^{-1}$ | $|A-B|$ km s$^{-1}$ kpc$^{-1}$ | Rotational velocity km s$^{-1}$ |
|---|---|---|---|---|
| Oort [26,27] | 19 | -24 | 33 | 247.5 |
| Kerr & Lynden-Bell [32] | 14.4 ± 1.2 | -12.0 ± 2.8 | 26.4 | 198.5 |
| Comeron et al. [33] | 12.9 ± 0.7 | -16.9 ± 1.1 | 29.8 | 223.5 |
| Feast a Whitelock [34] | 14.82 ± 0.84 | -12.37 ± 0.64 | 27.19 | 203.9 |
| Olling & Dehnen [35] | 15.9 ± 2 | -16.9 ± 2 | 32.8 | 246 |
| Branham, R. [36] | 16.08 ± 0.72 | -10.74 ± 0.65 | 26.78 | 200.8 |
| Branham, R. [37] | 14.85 ± 7.47 | -10.85 ± 6.83 | 25.43 | 190.7 |
| Bovy [38] | 15.3 ± 0.4 | -11.9 ± 0.4 | 27.2 | 204 |
| Chengdong et al [31] | 15.1 ± 0.1 | -13.4 ± 0.1 | 28.5 | 213.7 |
| This work | 15.6 ± 1.6 | -13.9 ± 1.8 | 29.5 | 221.2 |

## 5. Conclusion

In this work, we calculate the kinematical parameters and the Oort constants with a sample of 1444 red giants from SEGUE-1 and SEGUE-2 surveys. We divided the sample into three subsamples, the inner halo, the outer halo and the stars near the galactic plane. The velocity dispersions, projected distances and solar velocities for each subsample are computed. The derived velocity dispersions ($\sigma_1$, $\sigma_2$, $\sigma_3$) for the two halo components are as follows: Inner Halo = (265.86, 187.48, 138.29) km s$^{-1}$, Outer Halo = (286.01, 218.15, 174.36) km s$^{-1}$, and the stars near the galactic plane = (281.12, 190.22, 139.19) km s$^{-1}$. The solar velocities in km s$^{-1}$ are 213.4±14.6, 210.14±14.5 and 214.1±14.6 for the inner halo, outer halo, and galactic plane respectively.

We adopt the Oort constants, $A = 15.6 \pm 1.6$ km s$^{-1}$ kpc$^{-1}$ and $B = -13.9 \pm 1.8$ km s$^{-1}$ kpc$^{-1}$, the angular velocity $|A - B| = 29.5 \pm 0.2$ km s$^{-1}$ kpc$^{-1}$ implying a rotational velocity of $221.25 \pm 26.66$ km s$^{-1}$ assuming the distance to the Galactic center as 7.5 kpc. Our results indicate that the rotation curve $-(A+B)$ equals $-1.7 \pm 0.3$ km s$^{-1}$ kpc$^{-1}$, which indicates that the gradient of the rotation curve is negative and the circular velocity decreases in the galactic halo.

**Acknowledgments**

The authors gratefully acknowledge the approval and the support of this research study by the grant no. SCI-2018-3-9-F-7638 from the Deanship of Scientific Research at Northern Border University, Arar, Saudi Arabia.